\documentclass{article}
\usepackage[hscale=0.8, vscale=0.8]{geometry}
\usepackage{indentfirst}
\usepackage{amsmath}
\usepackage{graphicx}
\usepackage{subcaption}
\usepackage{caption}
\usepackage{setspace}
\usepackage{cite}
\usepackage{authblk}
\title{Theory and Practice of Electron Diffraction from Single Atoms and Extended Objects using an Electron Microscope Pixel Array Detector}
\author[1]{Michael C. Cao}
\author[1]{Yimo Han}
\author[1]{Zhen Chen}
\author[2]{Yi Jiang}
\author[3]{Kayla X. Nguyen}
\author[1]{Emrah Turgut}
\author[1,4]{Greg Fuchs}
\author[1,4]{David A. Muller}
\affil[1]{School of Applied and Engineering Physics, Cornell University, Ithaca, NY, USA}
\affil[2]{Department of Physics, Cornell University, Ithaca, NY, USA}
\affil[3]{Department of Chemistry and Chemical Biology, Cornell University, Ithaca, NY, USA}
\affil[4]{Kavli Institute for Nanoscale Science, Cornell University, Ithaca, NY, USA}
\begin{document}
\bibliographystyle{unsrt}
\maketitle
\begin{abstract}
What does the diffraction pattern from a single atom look like?  How does it differ from the scattering from long range potential?  With the development of new high-dynamic range pixel array detectors to measure the complete momentum distribution, these questions have immediate relevance for designing and understanding momentum-resolved imaging modes.  We explore the asymptotic limits of long range and short range potentials.  We use a simple quantum mechanical model to explain the general and asymptotic limits for the probability distribution in both and real and reciprocal space.  Features in the scattering potential much larger than the probe size cause the bright-field disk to deflect uniformly, while features much smaller than the probe size, instead of a deflection cause a redistribution of intensity within the bright-field disk.  Because long range and short range features are encoded differently in the diffraction pattern, it is possible to separate their contributions in differential phase contrast (DPC) or Center-of-Mass (CoM) imaging.  The shape profiles for atomic resolution CoM imaging are dominated by the shape of the probe gradient and not the highly-singular atomic potentials or their local fields.  Instead, only the peak height shows an atomic-number sensitivity, whose precise dependence is determined by the convergence angle. At lower convergence angles, the contrast oscillates with increasing atomic number, similar to bright field imaging.  The range of collection angles impacts DPC and CoM imaging differently, with CoM being more sensitive to the upper cutoff limit, while DPC is more sensitive to the lower cutoff.
\end{abstract}
\section{Introduction}
A new generation of high-speed, pixelated detectors has expanded the possibility for collecting full scattering information by recording diffraction patterns from a focused probe in a scanning transmission electron microscope (STEM), creating a rich and phase-sensitive four dimensional (4D) data set.  For high dynamic range detectors where the full diffraction pattern, including the central beam can be recorded without saturation, all traditional STEM imaging modes can be reconstructed quantitatively  and simultaneously, with their signals placed on an absolute scale \cite{Tate2016}. However, the rich information contained in these spatially-resolved diffraction patterns is well suited to more sophisticated imaging modes, some new, and some long-envisaged that until now have been handicapped by detector technology.

Shortly after the practical demonstration of field emission STEM instruments capable of forming an small probe\cite{Crewe1970}, researcher began to explore potential imaging modes that exploited the phase information encoded across the convergent beam diffraction patterns formed by a coherent, focused electron beam.  Rose\cite{Rose1974} considered the phase distribution across the bright field disk in the presence of aberrations as a tunable phase plate and proposed a series of ring-like detectors matched to regions of opposite phase to obtain a phase-contrast imaging with higher collection efficiency and resolution than a traditional bright field (BF) STEM image. As Rose noted in the paper, matching the aberration function to the detector and not having it drift would be challenging and it was not until actual phase plates and pixelated detectors were available, that a generalized version of this approach was implemented\cite{Ophus2016}.

To overcome the need to rely on aberrations for phase detection in STEM, Dekker and de Lang\cite{Dekkers1974} proposed another differential phase contrast (DPC) method using  a split detector divided into quadrants that essentially would measure the gradient of a weak phase object without the need for a phase plate, and showed optimal performance for an aberration-free, in-focus probe, was robust to contrast reversals with defocus, and with an information limit double that of traditional bright field STEM phase contrast image.   This is the same information limit for annular dark field (ADF) STEM, and as Rose later pointed out[6],  DPC (like ADF) has a contrast transfer function typical of a self-luminous object and incoherent imaging.  Rose also noted that the phase object itself could be obtained by integration, although the analogue integration schemes of the day would lead to large, low frequency noise instabilities\cite{Rose1977}.   With the advent of widely-available Fourier-based methods, integrated DPC (iDPC) was implemented first for x-ray microscopy\cite{Hornberger2007} and later electron microscopy\cite{Close2015,Lazic2016}.  Early applications of DPC in STEM included mapping magnetic fields and domain walls at medium resolution \cite{chapman1978,Waddell1977}. With the widespread availability of aberration correctors, DPC for atomic-resolution lattice imaging has also become practical \cite{Shibata2012,Muller2014}. 

The contrast mechanisms and scattering distributions are actually somewhat different for the medium resolution imaging of magnetic domains and atomic-resolution imaging, and the uniform deflections observed in the former case\cite{chapman1978}, are not seen in the latter\cite{Muller2014}, or for that matter at domain boundaries in ferroelectrics\cite{MacLaren2015}.  Our goal here is to rationalize these different results and provide a general picture of contrast changes across the diffraction pattern so as to guide development of new and optimized imaging modes with pixelated detectors.  Considerable early work has already been done, and with the exception of ptychography \cite{Hoppe1969,Nellist1995,Rodenburg1992} has largely focused on identifying simple detection schemes.  Nevertheless, there is still great value in these early derivations that can be extended to the more modern pixelated detectors.  For instance, Figure 1 of Dekkers and de Lang\cite{Dekkers1974} would be immediately recognizable to modern researchers performing single-side-band ptychography – both are targeting the same contrast changes  across the BF disk, and consequently both iDPC and BF ptychography show the same information limit \cite{Pennycook2015}.  However, as more pixels are added to the detector, the BF ptychography can construct a more optimal sampling of the disk overlaps, leading to improvements in contrast and collection efficiency\cite{Pennycook2015}. 

Similarly, using a pixelated detector also allows for improvements on DPC as a direct phase detection method.  Waddell and Chapman showed that  the gradient of a potential  is recovered exactly within the strong phase approximation by calculating the center of mass (CoM) of the diffraction pattern instead of the DPC signal\cite{Waddell1977,Waddell1979}. This CoM signal would of course require a pixelated detector to perform the linear weighting needed to obtain the first moment. The first moment or center-of-mass measurement is a better measurement of phase gradient compared to the split quadrant detector, at least in terms of uniformity of the contrast transfer function (CTF) \cite{Lazic2016,Waddell1979,Muller2017}. As noted more recently, the CoM signal has a physical meaning in its own right, independent of the strong phase approximation – it is the expectation value of the quantum mechanical probability current flow of the electron beam through the sample\cite{Muller2014,Lubk2015}.  This more modern interpretation is helpful in thinking of effects beyond the weak scattering limit, and while it strictly applies to CoM imaging, DPC is often a sufficiently close approximation that it is also applied there as well.

Both DPC and CoM images are formed by summing over the diffraction pattern using an anti-symmetric weighting function over the detector.  DPC uses a weighting function of 1 and -1 over the two halves of the detector.  CoM uses the coordinate in momentum space k ⃗ as the weighting function.  We will use this convention when referring to DPC or CoM images.  With these weighting functions, any asymmetry in the diffraction pattern would then lead to a difference signal.  Classically, this asymmetry was due to a small displacement of the bright-field disk.  However, experiments have measured a difference signal even when the disk is not displaced \cite{MacLaren2015}.  Instead, there is a redistribution of intensity within the disk, and this is commonly seen in multislice simulations of diffraction pattern mapping and DPC imaging at the atomic scale\cite{Close2015,Muller2014,Muller2017}.  Here we use a phase object approximation to separate these phenomenon as a result of different length scales.  By choosing the integration angles, we show how CoM images separate large from small features.  Furthermore, we highlight an important caveat when interpreting a CoM image when the feature size is small compared to the probe, cautioning how we should not interpret atomic CoM/DPC signals as mapping the shape of the potential, but rather the gradient of the probe shape instead.  We also consider the effect of detector geometry on coherence and optimizing the signal to noise ratio.  

We provide experimental examples using a pixelated detector with high electron sensitivity and a fast readout time.  While 4D diffraction data can be collected using a traditional CCD \cite{Kimoto2011}, CCDs are significantly limited by readout time and signal saturation. Further, early pixelated detectors were limited by their poor dynamic range or limited maximum dose before saturation to studying only the BF disk, or only the high-angle scattering but not both, and generally with an insufficient number of electrons/per pixel to discern the contrast changes of interest for this work.  To address these limitations, we apply our recently developed electron microscope pixel array detector (EMPAD) capable of single electron detection with a 140:1 signal to noise detection, 1,000,000:1 electron dynamic range, and a 1 kHz readout speed \cite{Tate2016}.  With this detector weak features inside the high intensity central disk can be resolved simultaneously with low-intensity details at large scattering angles, for a small electron beam placed on and between single atoms in two-dimensional materials -- useful and stable test objects for illustrating our main theoretical points.

\section{Center-of-Mass Contrast Changes in Momentum Space}
Classically, the bright-field disk shifts away from the unscattered position due to electric or magnetic fields uniformly deflecting the electron beam.  However, despite the presence of a field, there are numerous instances where the boundaries of the central disk do not move.  Instead, there is a redistribution of intensity within the bright-field disk itself.  By using a quantum mechanical approach and treating the sample as a phase object, we develop a simple model that explains both behaviors as a result of differing length scales between probe and scattering potential.

In order to calculate a diffraction pattern, we start by writing the probe wavefunction in terms of the probe-forming aperture and the angular aberrations as
\begin{equation}
\Psi_0(\vec{r}) = \dfrac{1}{2\pi}\int A(\vec{k})\exp\left(i\chi(\vec{k})\right)\exp\left(i\vec{k}\cdot\vec{r}\right)d\vec{k},
\end{equation}
where $\chi(\vec{k})$ are the aberrations in the lens and impart a phase term to the incoming wave.  For convenience we will assume an aberration free probe throughout this work unless otherwise stated.  This simplifies the interpretation, but does not limit the generality.  The $A(\vec{k})$ term is the aperture function and is defined as:
\begin{equation}
\label{Aperture}
A(\vec{k}) = \left\{
\begin{aligned}
1 &\quad k\leq k_0\\
0 &\quad k>k_0
\end{aligned}
\right.,
\end{equation}
where $k_0$ is the maximum angle of the aperture.  After the probe is formed, the STEM rasters the probe over the sample.  We model the interaction with the sample using the strong phase approximation, so that after interaction, the probe has acquired an additional phase:
\begin{equation}
\Psi(\vec{r},\vec{r}_p) = \Psi_0(\vec{r}-\vec{r}_p)\exp\left(i\sigma V(\vec{r})\right),
\end{equation}
where $\vec{r}_p$ is the scanning position and σ is the interaction parameter \cite{Kirkland2010}.  The validity of the strong phase approximation is dependent on sample composition and thickness.  Simulations of GaN by Muller-Caspery et al. recommend a thickness under a few nanometers \cite{Muller2017}.  The diffraction pattern is collected by a detector placed in the back focal plane.  The diffraction pattern is related to the Fourier transform of our exit wavefunction:
\begin{equation}
I(\vec{k}) = \left|\mathcal{F}\left[\Psi(\vec{r},\vec{r}_p)\right]\right|^2
\end{equation}
\begin{figure}
	\centering
	\includegraphics[]{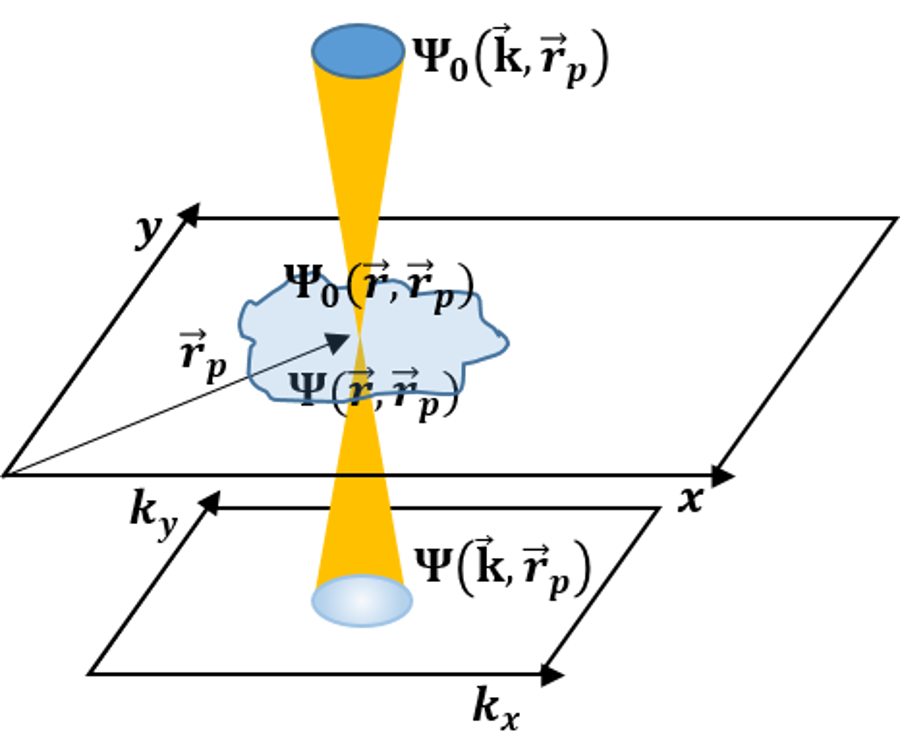}
	\caption{Schematic of diffraction pattern formation in the kx-ky plane as probe $\Psi_0$ centered at point $\vec{r}_p$ is scanned in the x-y plane by shifting $\vec{r}_p$.}
	\label{Schematic}
\end{figure}
This process is outlined in Figure \ref{Schematic}.  Now we look at the asymptotic cases between the sample potential $V(\vec{r})$ and probe size. 
\subsection{Probe Size \textless\textless\space Feature Size}
When the probe size is much smaller than the feature size, we can model the sample potential as a linear ramp $V(\vec{r}) = E_0 x$.  In the strong phase approximation, our initial wavefunction picks up a phase proportional to the strength of the sample potential to form our exit wavefunction:
\begin{equation}
\Psi(\vec{r}, \vec{r}_p) = \Psi_0(\vec{r} - \vec{r}_p)\exp\left(i\sigma E_0 x\right).
\end{equation}
Taking the Fourier transform and squaring the amplitude gives the diffraction pattern:
\begin{equation}
\left|\Psi(\vec{k}, \vec{r}_p)\right|^2 = \left|\Psi_0(\vec{k} - \sigma E_0\hat{x})\right|^2.
\end{equation}
\begin{figure}
	\centering
	\includegraphics[scale=0.75]{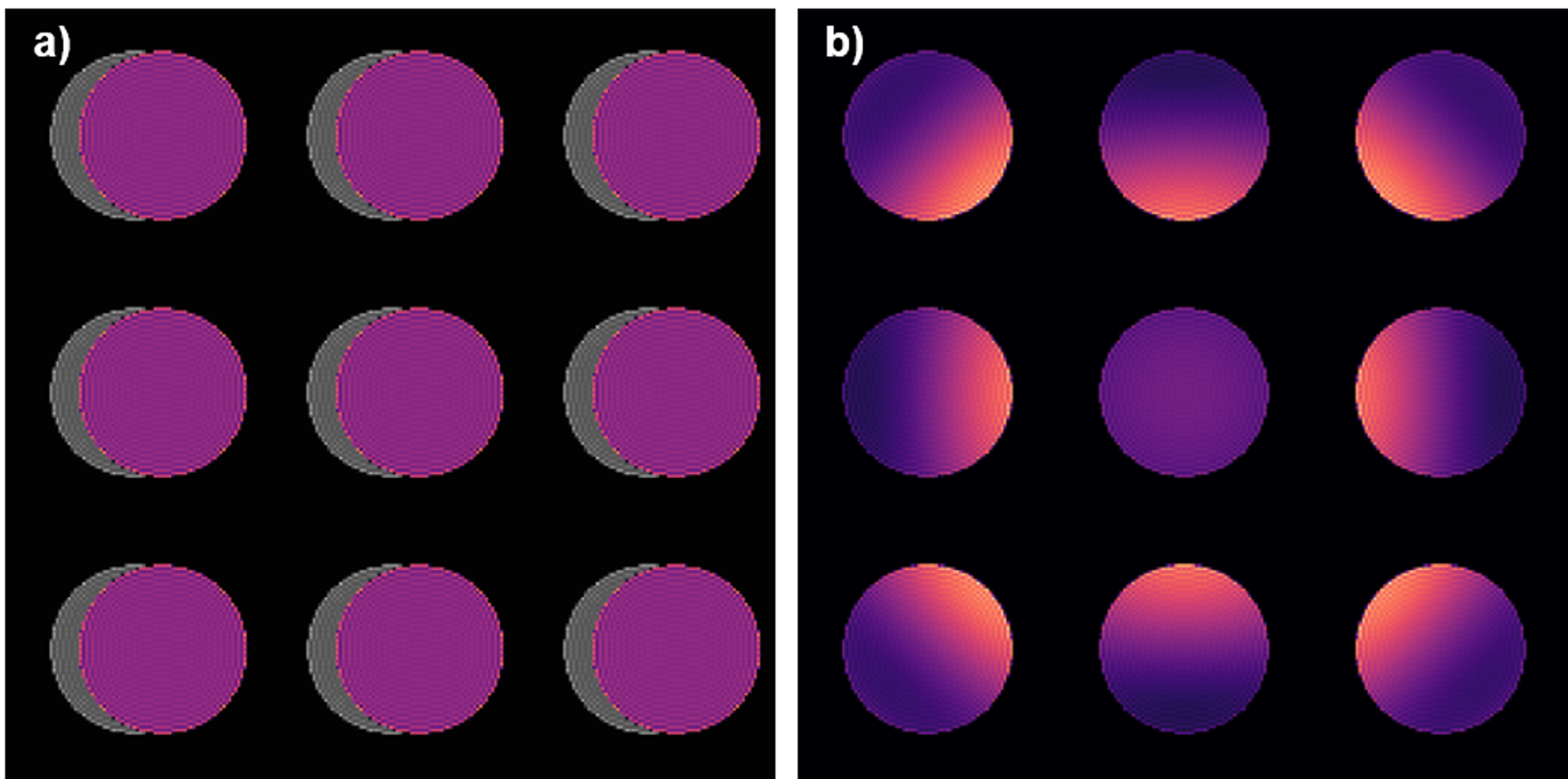}
	\caption{Diffraction patterns for scanning points 0.25 Å apart with a 30 mrad aperture for a) a constant field that causes the disk to shift by 10 mrad and b) an atomic-sized potential field, which instead of a shift, causes an asymmetry in the intensity within the disk.  (a) and (b) describe the limiting cases where the potential is much larger, and much smaller than the probe size respectively.}
	\label{Diffraction Patterns}
\end{figure}
The resulting diffraction pattern is just the original bright-field disk of the unscattered beam uniformly shifted to the right as shown in Figure \ref{Diffraction Patterns}a.  This shift is proportional to the strength of the field.  The result is the same as the classical result, and the shift from the field shifts the entire diffraction pattern uniformly.  For small shifts, the asymmetry in placement creates a difference signal on a split quadrant detector that is also proportional to the field strength.
\subsection{Probe Size \textgreater\textgreater\space Feature Size}
In the other limiting case where the feature size is much smaller than the probe, we can model the sample potential as a delta function $V(\vec{r})=V\_0 \delta(\vec{r})$.  This time, for analytic simplicity we use a weak phase approximation model for our exit wavefunction:
\begin{equation}
\Psi(\vec{r}, \vec{r}_p) = \Psi_0(\vec{r} - \vec{r}_p)[1 + i\sigma V_0\delta(\vec{r})].
\end{equation}
The expression for the diffraction pattern is then
\begin{equation}
\label{DeltaPotentialDP}
\left|\Psi(\vec{k},\vec{r}_p)\right|^2 = \left|A(\vec{k})\right|^2 - 4\pi A(\vec{k})\sigma V_0 k_0\dfrac{J_1\left(k_0\left|\vec{r}_p\right|\right)}{\left|\vec{r}_p\right|}\sin(\vec{k}\cdot\vec{r}_p) + \left|2\pi\sigma V_0 k_0 \dfrac{J_1\left(k_0\left|\vec{r}_p\right|\right)}{\left|\vec{r}_p\right|}\right|^2.
\end{equation}
The first term is just the initial probe’s bright-field disk, but the second term gives structure to the bright-field disk.  The third term is 2nd order in the scattering and so much weaker in intensity.  For the delta function potential, this term is uniform in k, providing uniform offset to the diffraction pattern that changes intensity as the probe is scanned across the potential, but will not contribute to a center of mass signal.  It will contribute to the annular dark field signal and with a more realistic potential will become the incoherent convolution of the probe with the square of the potential.

Instead, the first order change in contrast in the diffraction pattern is dominated by the 2nd term -- $V_0  sin⁡(\vec{k}\cdot[\vec{r}_p])$ in particular.  Instead of an asymmetry in the placement of the disk seen in the other regime, the k-dependence causes the asymmetry to occur in the intensity of the bright-field disk itself and is probe position dependent as shown in Figure \ref{Diffraction Patterns}b.  The modulation is proportional to $V_0$ and not $V_0^2$ as in ADF imaging, so as we will see below, the atomic number (Z) dependence for CoM and DPC imaging is closer to Z than $Z^2$.  It’s important to note that the presence of a single $A(\vec{k})$ in the second term ensures that no intensity redistributions outside the central disk are allowed, so shifts of the disk boundary as seen in the classical case cannot occur here.

\begin{figure}
	\centering
	\includegraphics[]{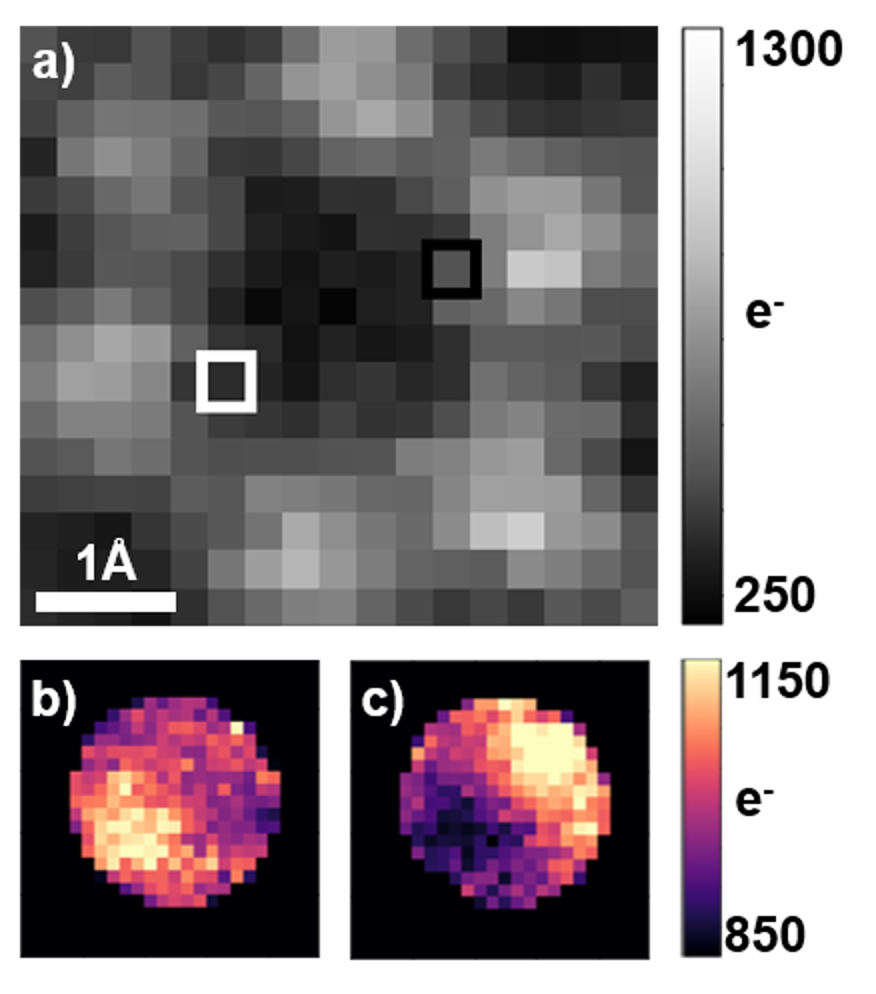}
	\caption{a) Reconstructed HAADF image of WSe$_2$ from diffraction data collected on the EMPAD at 80 keV with a 21.4 mrad aperture.  White and black boxes outline the pixels associated with diffraction patterns shown in b) and c) respectively.  Both diffraction patterns show the redistribution in intensity due to the probability current flow towards the nearest nuclei, i.e. the probe to the right of an atom shows enhanced probability current flow to the left, and vise-versa.}
	\label{WSe2 Diffraction Pattern}
\end{figure}

The intensity redistribution is also observable experimentally –- as shown for the experimental diffraction data collected on a monolayer WSe2 sample shown in Figure \ref{WSe2 Diffraction Pattern}.  The data was collected on the Cornell-developed EMPAD using an 80 keV electron beam focused to atomic dimensions using a 21.4 mrad aperture.  Figure \ref{WSe2 Diffraction Pattern}b,c shows the redistribution of intensity when the probe places close to nearby atoms in the monolayer lattice.

Consequently, whether the CoM/DPC signal is caused by a shift of the bright disk or intensity redistribution inside the disk is dependent on feature and probe size.  Furthermore, short range and long range field information is encoded in separate ways.  Low spatial frequency information about long-range potentials is found in the uniform displacement of the bright-field disk, but high spatial frequency information about short-range potentials is found in the redistribution of intensity within the disk itself.   

\subsection{Separating Long-Range and Short Range Potentials in Center-of-Mass Imaging}
The difference in how information about the long-range and short range potentials are encoded in diffraction space has been applied empirically to enhance magnetic contrast.  Chapman et al used an annular DPC detector with an inner angle that is very close to the semi-angle of the bright-field disk in order to emphasize the contrast from the signal that shifts the bright-field disk – in their case a slowly varying magnetic field \cite{chapman1990}.  Similarly, Krajnak et al attempted to isolate only the shift of the bright disk from the intensity redistribution in post-processing again in order to enhance magnetic contrast and suppress non-magnetic features \cite{Krajnak2016}.  Using our model for the diffraction pattern formation, we understand that by placing the inner angle at the edge of the unscattered beam, we significantly filter out most of the information from short range fields that are contained only within the bright disk, and distributed across the disk.  However, long range fields are easily detected because even the most minimal shift will move the beam outside the inner detector angle and create a difference signal.  The question as to why this worked to separate magnetic contrast from grain contrast has less to do with magnetic fields fundamentally always shifting the bright field disk than in many materials magnetic potentials vary slowly.  Provided the electron beam shape is kept much smaller than the thickness of any domain wall, the magnetic potential will appear to be slowly varying, thus meeting the conditions for shifting the bright-field disk.  Grain boundaries tend to be relatively compact, with structural distortions rarely extending more than a few nm, so for a large probe needed to detect small angular deflections, grain boundaries (and sometimes grain size in thin films) can be smaller than probe size.  However, it is important to realize this filtering is essentially a smoothing or low pass filtering of the image rather than a general selection for magnetism.  This can be made more explicit by considering the phase gradient transfer function for DPC imaging calculated by Majert and Kohl for an annular detector.  When the annulus is reduced to barely overlapping the unscattered beam, the transfer function has its most extreme low-pass filter profile \cite{Majert2015}.

From the diffraction pattern model of section 2.2, we see that all the high frequency data is contained within the bright disk.  Conversely, by reducing the outer angle on the detector to inside the bright-field disk, the high frequency contrast is improved by suppressing the low-frequency contrast contained near the edges of the disk.  Because the shift of the disk caused by the slowly varying field is typically very small, it is usually unnecessary to re-center the detector.  However, if the shift is significant, it would be necessary to re-center to isolate the high frequency portion of the signal. 

\begin{figure}
	\centering
	\includegraphics[]{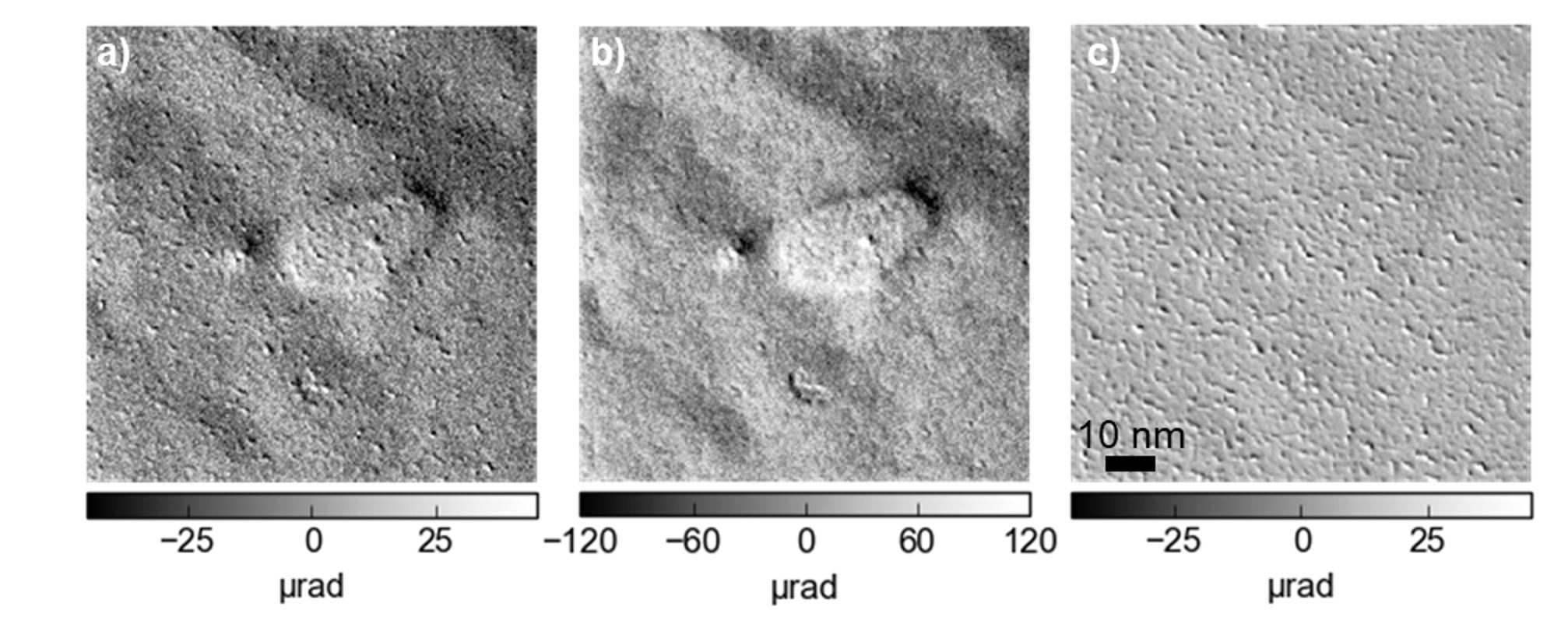}
	\caption{COMx images of ferromagnetic polycrystalline-FeGe.  a) COMx image formed by integrating over all angles show magnetic domains and grain contrast.  b) By only integrating over an annulus around the edge of the bright-field disk, the domains are displayed more clearly.  c) By integrating inside the disk, the grain contrast is shown while the magnetic domain contrast is greatly suppressed.}
	\label{FeGe}
\end{figure}

Figure \ref{FeGe} demonstrates the long range/short contrast filtering with a thin-film sputtered polycrystalline ferromagnetic phase of FeGex imaged at 100K.  Integrating all angles for the CoM image shows some of the magnetic domain contrast, but by only integrating close to the edge of the bright disk, the magnetic domains show up more clearly.  On the other hand, integrating angles within the bright disk picks up the grain contrast without showing the magnetic domains.  The effectiveness of this strategy relies on the probe size being larger than the grain contrast, but smaller than the magnetic field variation.  As the probe size is reduced to improve spatial resolution, the smaller the spatial range of grain contrast that can be reduced as well.

\section{Center-of-Mass Contrast in Real Space}
\begin{figure}
	\centering
	\includegraphics[]{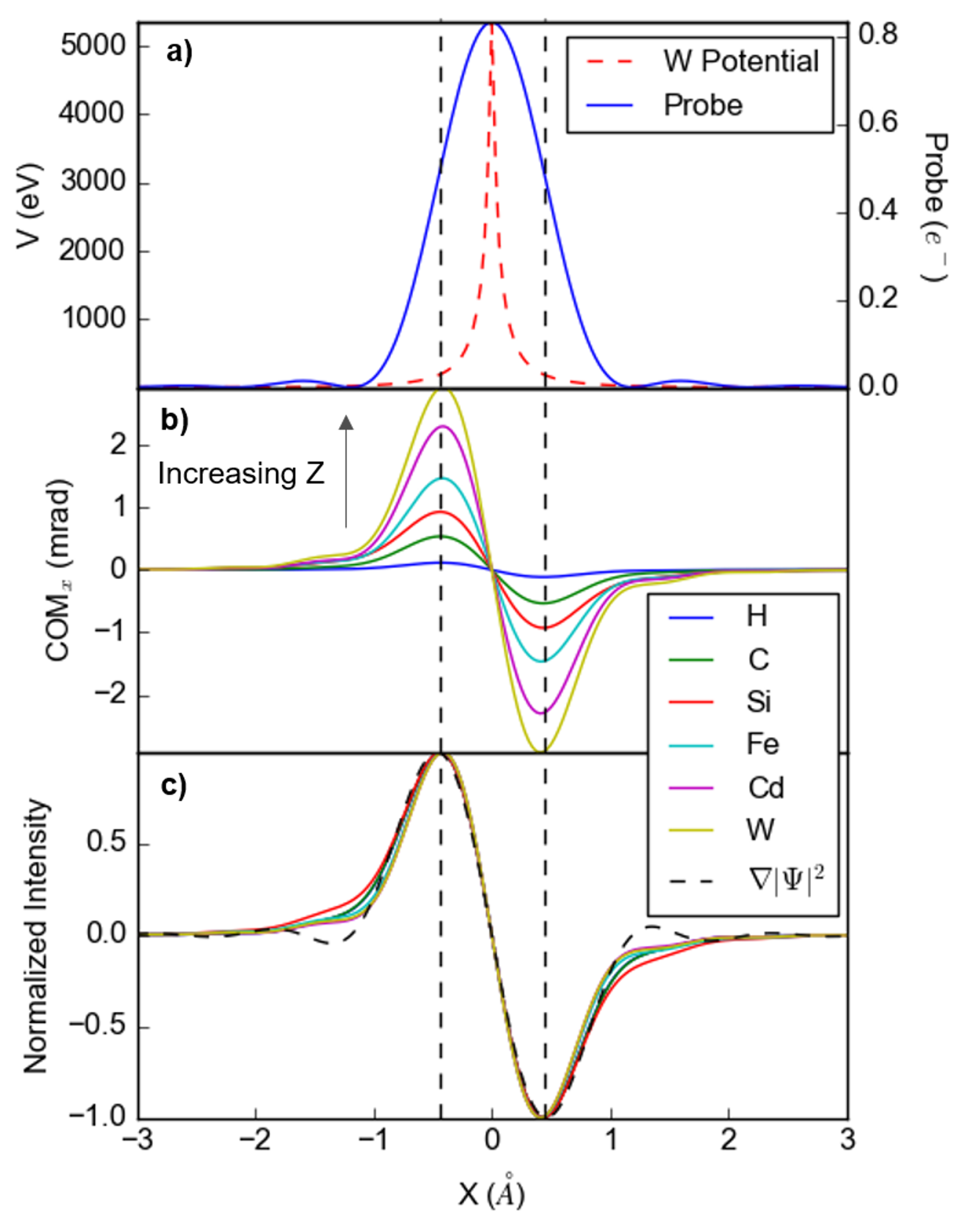}
	\caption{COMx images of single atoms.  A) Simulated non-aberrated 80 keV probe with a 21.4 mrad aperture has a significantly wider profile than the potential from a single tungsten atom.   b) Simulated COMx profiles of light and heavy atoms show difference in height, but the normalized plots in c) show that the profile shapes are very similar, and closely follow the gradient of the probe profile – both of which peak at the inflection points of the initial probe.}
	\label{Atomic COM}
\end{figure}
We can also examine the CoM image itself in real space.  Previous works have connected the CoM to the probability current and to local electric fields \cite{Lazic2016,Muller2014,Lubk2015}.  The first moment or center-of-mass image has each pixel value equal to the centroid of the diffraction pattern:
\begin{equation}
\label{COMDef}
CoM(\vec{r}_p) = \int \vec{k}\left|\Psi(\vec{k}, \vec{r}_p)\right|^2 d\vec{k}.
\end{equation}
A simple commutation relates this expression to the expectation value of the momentum operator:
\begin{equation}
CoM(\vec{r}_p) = \int \Psi^* (\vec{k}, \vec{r}_p)\vec{k}\Psi(\vec{k},\vec{r}_p)d\vec{k} = \dfrac{<\vec{p}>}{\hbar}.
\end{equation}
Further calculation \cite{Muller2014,Lubk2015} shows its relation to the 2D probability current in the specimen plane:
\begin{equation}
\label{GeneralCOMResult}
CoM(\vec{r}_p) = \dfrac{1}{2i}\int\Psi^*(\vec{r}, \vec{r}_p)\vec{\nabla}\Psi(\vec{r},\vec{r}_p) - \Psi(\vec{r}, \vec{r}_p)\vec{\nabla}\Psi^*(\vec{r},\vec{r}_p)d\vec{r},
\end{equation}
\begin{equation}
CoM(\vec{r}_p) = \dfrac{m}{\hbar}\vec{j}(\vec{r}_p).
\end{equation}
This form is exact for all sample thicknesses.  Additionally, by using a strong phase approximation, we can connect the CoM image to the scattering potential to get another expression for thin samples:
\begin{equation}
CoM(\vec{r}_p) = i\int\Psi^*_0(\vec{r} - \vec{r}_p)\vec{\nabla}\Psi_0(\vec{r} - \vec{r}_p)d\vec{r} + \sigma\int\vec{\nabla}V(\vec{r})\left|\Psi_0(\vec{r} - \vec{r}_p)\right|^2 d\vec{r}.
\end{equation}
If the beam is symmetric, then the first term goes to zero, otherwise there is a constant background, and the second term is a cross correlation\cite{Lazic2016} between the potential gradient and the probe:
\begin{equation}
\label{SPOCOMRes}
CoM(\vec{r}_p) = \sigma\vec{\nabla}V(\vec{r}_p)\star\left|\Psi_0(\vec{r}_p)\right|^2.
\end{equation}

Because there is a cross correlation (or convolution for symmetric probes) between the probe and the potential gradient, we must be cautious when interpreting the image, especially in the regime where the feature size is small.  In the extreme limiting case where the potential is reduced to a delta potential $V_0 \delta(\vec{r})$, this leads to a CoM image of
\begin{equation}
\label{DeltaPotentialCOM}
CoM(\vec{r}_p) = -\sigma V_0 \vec{\nabla}\left[\left|\Psi_0(-\vec{r}_p)\right|^2\right].
\end{equation}
This is consistent with taking the first moment of eq. (\ref{DeltaPotentialDP}).  Notably, while the signal strength scales linearly with scattering strength of the sample, $\sigma V_0$, the spatial information in the image is related to the probe and not the potential.

This is shown in Figure \ref{Atomic COM} with a simulated non-aberrated probe at 80 keV with a 21.4 mrad aperture for light and heavy atoms.  All simulated potentials are stationary with no Debye-Waller factor included.  Thermal effects on CoM are less than a few percent as shown by Muller-Caspary et al. \cite{Muller2017}.  Since the probe width is much larger than the width of the simulated tungsten potential, the resulting CoMx linescan follows the probe profile rather than the atomic profile.    As seen in Figure \ref{Atomic COM}b,c this holds for both light and heavy atoms.  The main difference is the relative peak height, which is set by the Z scaling of the atomic potential.  By normalizing by height, we see the profiles are very similar, independent of Z.  While there are some weak features, these can be muddled by probe tails, residual aberrations and very small (sub-nm) changes in defocus in actual experiments.
\begin{figure}
	\centering
	\includegraphics[]{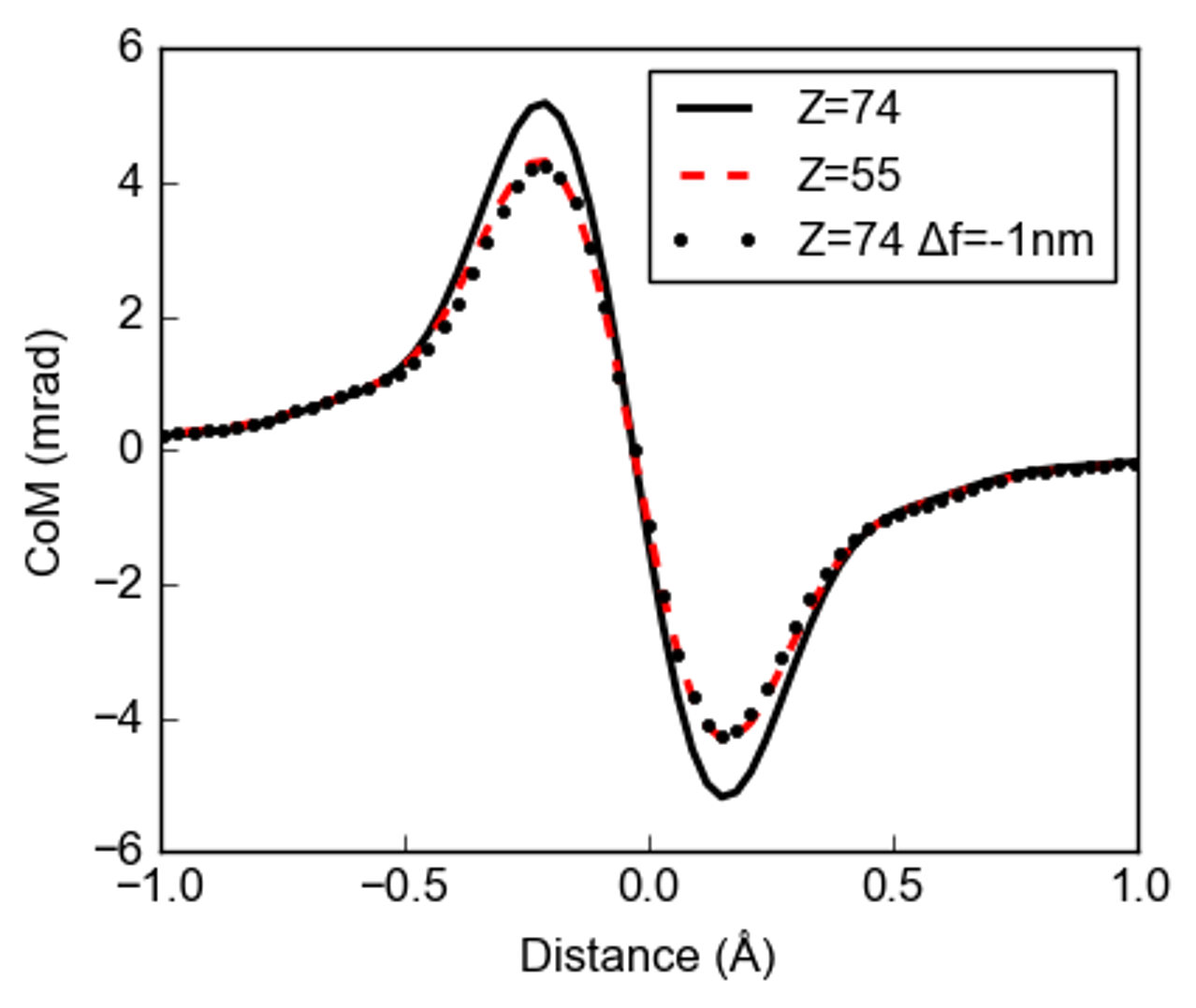}
	\caption{The effect of defocus on COM linescans across atomic potentials for a 300 keV aberration-free electron probe, with a 30 mrad probe-forming aperture.  At zero defocus, the tungsten (Z=74) and cesium (Z=55) profiles are distinguishable by their peak deflections.  However, with only a 1 nm defocus, the tungsten profile almost exactly matches the cesium profile.  In general, variations from defocus and other higher order aberrations much smaller than the alignment tolerances of the microscope give larger effects than charge transfers or bonding changes.}
	\label{Defocus COM}
\end{figure}
In Figure \ref{Defocus COM}, a 1 nm change in defocus produces a shift in mrad comparable to a change in atomic number by ∆Z=19.  (For comparison, a 1 nm defocus change is much smaller than the depth of field or defocus spread from chromatic blur which are typically 5nm or more). Similarly in Figure \ref{CC COM}, linescans of a tungsten potential with defocus spread with a standard deviation up to 3 nm due to chromatic aberration changes the peak height, but not the tails.  Contrast transfer functions (CTF) for integrated CoM (iCoM) and integrated DPC (iDPC) by Lazić et al. show defocus affects the lower to mid frequencies, consistent with our result \cite{Lazic2016}.  CoM and iCoM are linearly related, so their CTFs identical.  Rather than probing the shape of the potential (or details of the sample’s electric field), the shape of the signal maps the derivative of the probe shape, and is more sensitive to changes in the probe shape than the underlying potential.  While small changes to defocus only affected the peak height, other aberrations can have a stronger effect on the probe tails which will be reflected in the tails of the CoM image.  Overall, this cautions against any interpretation that we are mapping the atomic structure or bond charge distributions with an atomic CoM/DPC signal.

\begin{figure}
	\centering
	\includegraphics[]{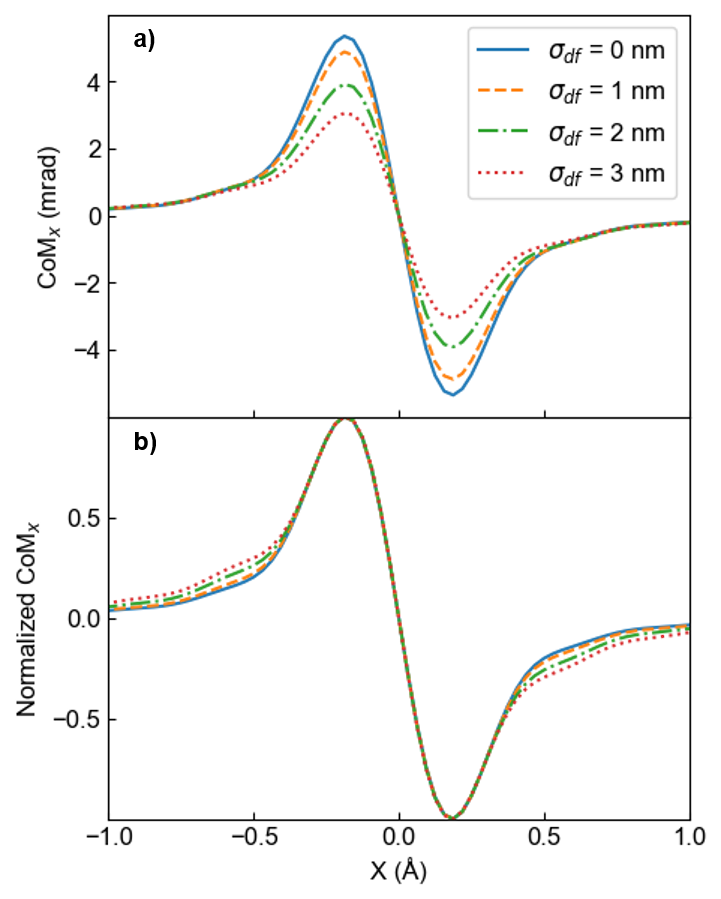}
	\caption{The effect of defocus spread due to chromatic aberration on a COMx linescan of a tungsten potential at 300 keV with a 30 mrad aperture.  Defocus spread values shown are the standard deviations.  a) Similar to figure 6, defocus spread lowers the peak height with little effect on the tails as if changing the atomic potential.  b)  Normalizing the heights shows a similar effect to figure 5c.}
	\label{CC COM}
\end{figure}
\begin{figure}
	\centering
	\includegraphics[]{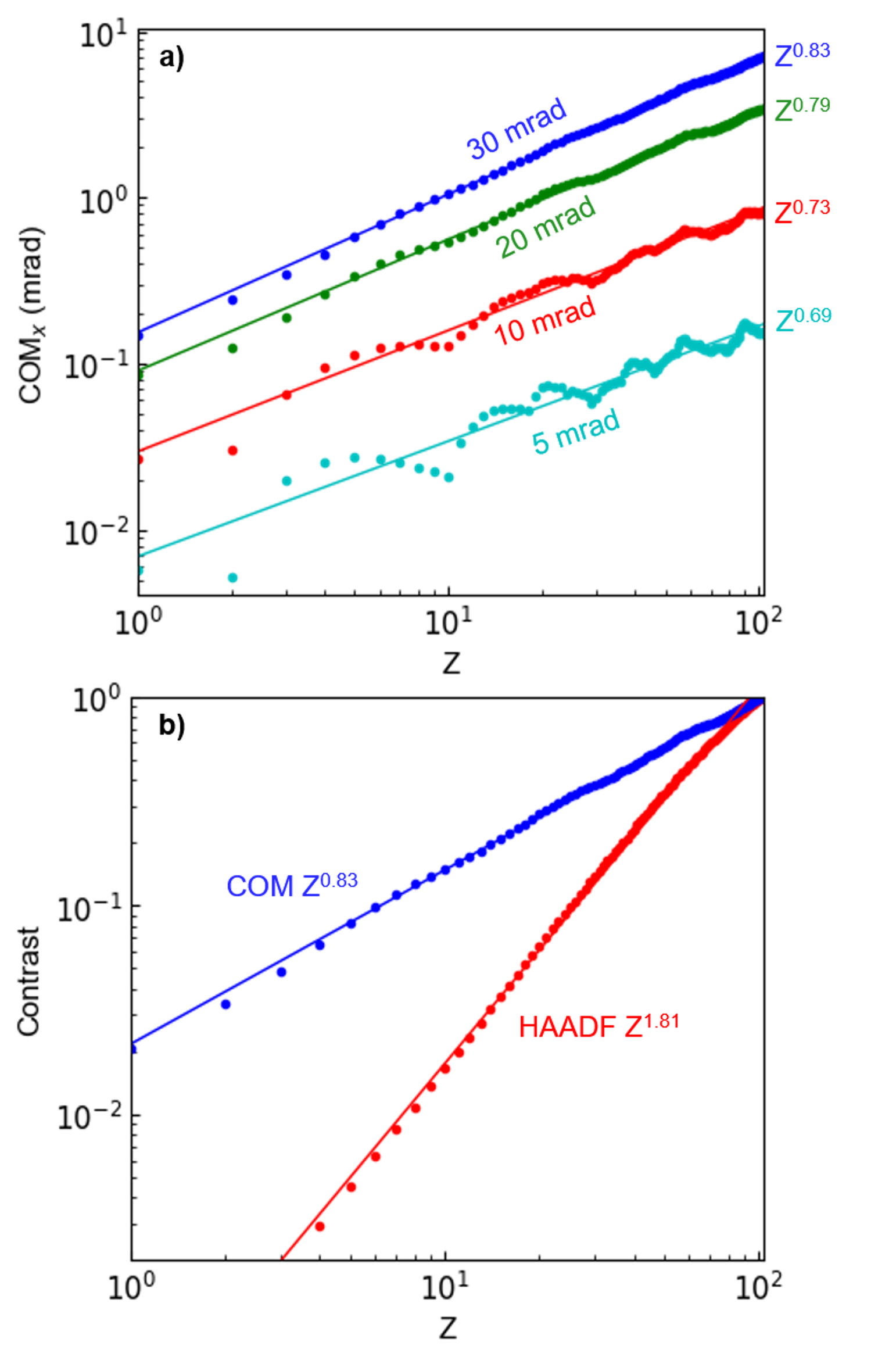}
	\caption{a) Atomic-Number dependence for the peak COM shift from a single atom as a function of probe-forming aperture semi-angle for a 300 keV diffraction-limited electron beam.  For small semi-angles, outer-shell valence trends dominate the Z-dependence, while for large semi-angles, the more monotonic nuclear contribution dominates.  This is reflected in the increasing exponent for the fitted power law with increasing aperture size.  b) Comparison with HAADF imaging shows that COM has a weaker Z dependence, making it easier to image light atoms in the presence of heavy atoms (300 keV, 30 mrad.  HAADF signal from a 3x-5x aperture size).}
	\label{Z Contrast}
\end{figure}
The relative heights of the CoM peaks can be used to measure the CoM contrast vs Z number. We compare its  Z-dependence with HAADF using simulations in Figure \ref{Z Contrast}.  CoM has less contrast between light and heavy atoms compared to HAADF.  This is why its cousin, the integrated CoM (iCoM) \cite{Lazic2016} is an ideal imaging mode for imaging samples with both light and heavy atoms.  Since iCoM is just a linear integration of CoM, the contrast scaling is the same for both.

The precise Z-dependence for CoM and DPC imaging depends on the convergence semi-angle and beam voltage, or more simply the cutoff measured in inverse Ångstroms.  For small cutoffs, the CoM contrast also does not monotonically increase with Z but shows periodic oscillations across the periodic table in much the same way that bright-field imaging does \cite{Kirkland2010}, indicating sensitivity to the shell structure of the atom.  When the cutoff angle is decreased, the probe becomes larger in real space and the atomic potential can be better approximated as a delta potential $V_0 \delta(\vec{r})$.  The resulting CoM profile is given in eq. (\ref{DeltaPotentialCOM}) where the peak height is determined by $V_0$, the mean inner potential, and structural information about the bonding of the atomic potential is lost and not reflected in the line profile.  As the cutoff angle is increased, the nuclear terms becomes more dominant, and the Z-dependence becomes smoother.

We can get a sense of these trends by Fourier transforming eq. (\ref{SPOCOMRes}) and noting that $\left|\Psi_0 (-\vec{r}_p)\right|^2$ is the point spread function, whose transfer is the contrast transfer function $\text{CTF}(\vec{k})$ to get
\begin{equation}
\label{CTF}
CoM(\vec{k}) = 2\pi i\sigma\vec{k}V(\vec{k})\text{CTF}(\vec{k}).
\end{equation}
For all focus and aberration values, the boundary points are fixed at $\text{CTF}(0) = 1$ and $\text{CTF}(k_0) = 0$.  For an in-focus aberration-free probe the CTF is a smooth ramp-like function between them that can be approximated as $\text{CTF}(k)\approx 1-k / k_0$  resulting in a bandpass sampling of the potential from 0 to $k_0$.  From eq. (\ref{CTF}) we see that for small cutoffs, $k_0$, we sample the small angle scattering which displays the electronic shell structure. As the cutoff $k_0$  becomes larger than the Thomas-Fermi angle, the contribution from the bare Coulomb potential becomes more heavily weighted and the Z-dependence becomes more linear.  DPC imaging shows a similar Z dependence to CoM imaging, with a Z dependent exponent just a few percent less.
\section{Effect of Detector Geometry}
\subsection{Coherence}
The question of signal to noise and coherency are additional considerations for the detector geometry.  If the angular range selected for the detector annulus is too small, it may not satisfy the condition for incoherent imaging.  Additionally, our diffraction pattern model in eq. (\ref{DeltaPotentialDP}) shows that the high frequency information is a small fluctuation of intensity within the larger flat background of the bright disk, whose poor signal to background ratio implies also a poor signal to noise.  Coherence is also affected by source size and the probe aperture, but we will focus on detector effects.  With these considerations, we start to explore the effects of the detector geometry.

We begin with the definition of the center-of-mass image in eq. (\ref{COMDef}), but we add a detector function $D(\vec{k})$:
\begin{equation}
CoM(\vec{r}_p) = \int\vec{k}\left|\Psi(\vec{k},\vec{r}_p)\right|^2 D(\vec{k})d\vec{k}.
\end{equation}
The detector function is defined similarly to our aperture function in eq. (\ref{Aperture}) but with some maximum detector angle $k_d$.  This calculation can be easily adapted for an annulus detector by repeating the calculation for the minimum angle and subtracting.  The final expression is
\begin{equation}
\label{Coherence}
CoM(\vec{r}_p) = \int\Psi(\vec{r}'',\vec{r}_p)\Psi^*(\vec{r}',\vec{r}_p)f(\vec{r}' - \vec{r}'')d\vec{r}''d\vec{r}'.
\end{equation}
where $f(\vec{r})$ is:
\begin{equation}
f(\vec{r}) = -\dfrac{i}{2\pi}\vec{\nabla}\dfrac{k_dJ_1(k_dr)}{r} = \dfrac{ik_d^2\vec{r}J_2(k_dr)}{2\pi r^2}.
\end{equation}
At the asymptotic limits, this function can be expressed as:
\begin{equation}
f(\vec{r}) = \left\{
\begin{aligned}
-i\vec{\nabla}\delta(\vec{r}) &\quad k_d\rightarrow\infty\\
\dfrac{ik_d^4\vec{r}}{16\pi} &\quad k_d\rightarrow 0
\end{aligned}
\right..
\end{equation}
Eq. (\ref{Coherence}) makes it clear that the function $f(\vec{r})$ is a coherency function.  

If this is taken at the asymptotic limit where $\vec{k}_d\rightarrow\infty$ with the strong phase approximation, this will reduce to eq. (\ref{GeneralCOMResult}) which is fully incoherent, consistent with the previous calculation.  At the other limit, we obtain the expression:
\begin{equation}
CoM(\vec{r}_p) = \dfrac{k_d^4\pi}{4}\dfrac{\partial}{\partial \vec{k}}\left[\left|\Psi(\vec{k},\vec{r}_p)\right|^2\right]_{\vec{k} = 0}.
\end{equation}
If this is applied to the diffraction pattern for a delta potential in eq. (\ref{DeltaPotentialDP}), the result is
\begin{equation}
CoM(\vec{r}_p) = -\dfrac{k_d^4\pi^2\sigma V_0 \vec{r}_p}{\left|\vec{r}_p\right|}J_1(k_0\left|\vec{r}_p\right|),
\end{equation}
which is shaped by the probe wavefunction, not its intensity, and scales linearly with the potential height, consistent with a coherent image.  The coherence length and cross-over from coherent to incoherent imaging is set by $k_0 / k_d$  and the collection angles need to be much smaller (at least 3x) than the convergence angle to see coherent effects.
\subsection{Signal to Noise Ratio}
\begin{figure}
	\centering
	\includegraphics[]{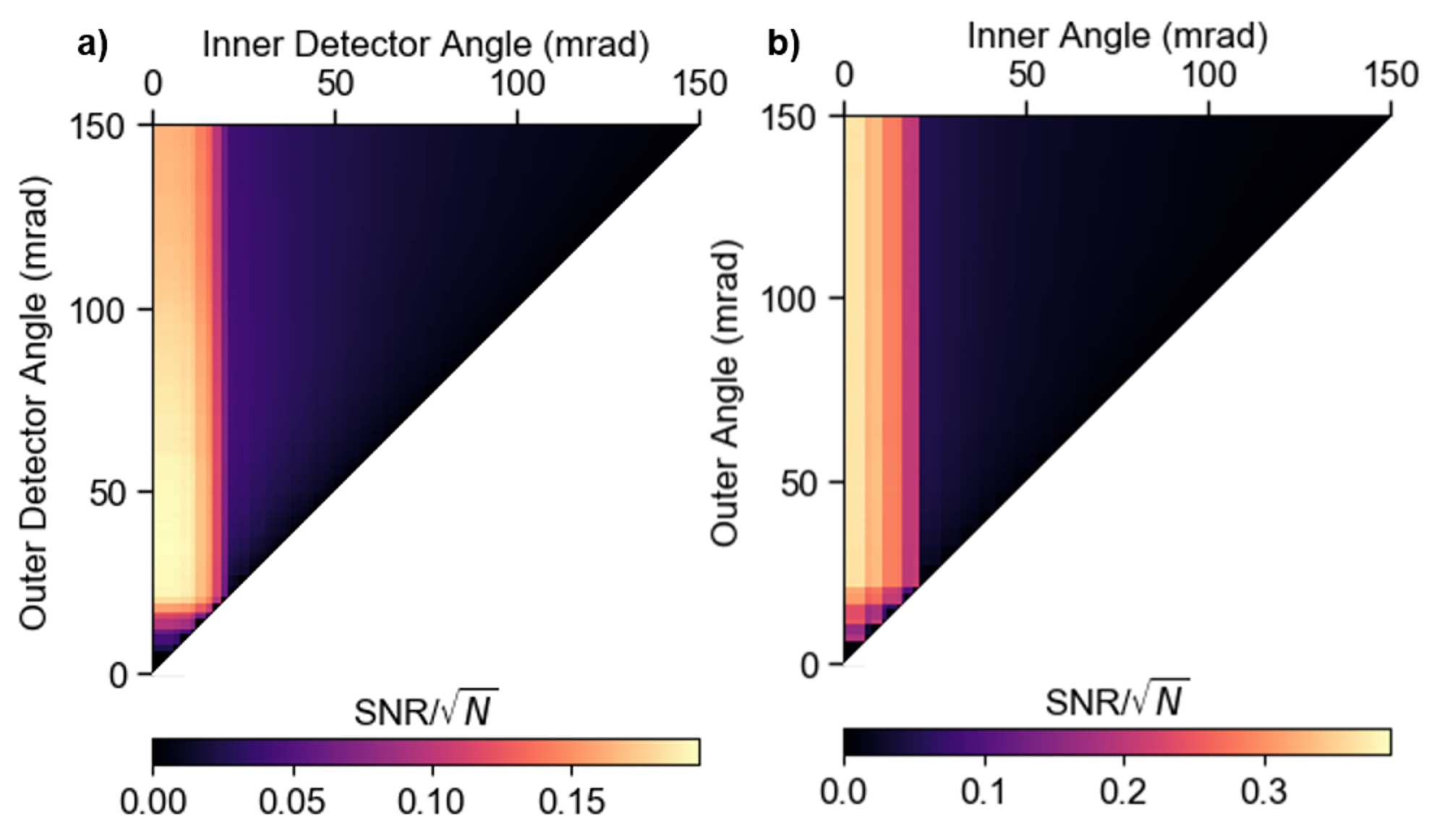}
	\caption{Simulated signal to noise ratio for a) COMx and b) DPCx image of a single tungsten atom with a 21.4 mrad aperture at 80 keV.  Both images show significant gains when integrating the edge of the bright disk.  However, the COM image SNR is reduced when collecting higher angles beyond ~2x the edge of the disk due to heavily weighing high angles with low electron counts.  Conversely, while the DPC image is not adversely affected by increasing the outer angle, it has a greater sensitivity to the inner angle cutoff.}
	\label{SNR}
\end{figure}
Another consideration is how the detector geometry affects the signal to noise ratio (SNR).  We simulated a single tungsten atom and found the scanning points that have the largest CoM signal.  The diffraction pattern was put through a poisson noise filter to simulate a shot-noise-limited signal.  This was repeated several times to obtain a robust statistical distribution.  The signal to noise ratio (SNR) is the true value of the signal divided by the standard deviation caused by the poisson noise.  This is then normalized by the incident electron counts N.  The simulation was done for a 80 keV beam with a 21.4 mrad aperture.

The results of these simulations are shown in Figure \ref{SNR} for both CoM and DPC imaging modes.  The best signal to noise was achieved with a 4 mrad to 33 mrad detector geometry for the CoM image, and 0 mrad to 47 mrad for DPC.  Most of the signal for both images come from the edge of the bright disk.  However, the CoM image loses SNR when expanding the outer angle since it is heavily weighting large angles which generally have very low electron counts.  On the other hand, the DPC image is largely outer angle independent once it is larger than the bright field disk, but more sensitive to inner angle since it weighs the small angle electron counts more heavily than does CoM.

Notably, the SNR for both DPC and CoM continue to increase for collection angles beyond the bright field disk (21.4 mrad) despite the weak phase approximation in eq. (\ref{DeltaPotentialDP}) showing information is contained only in the bright disk.  This is due to higher order terms outside the bright disk that also contribute to the CoM signal, which is shown in Appendix C.  Overall, the DPC weighting has a slightly higher SNR than the CoM weighting.  This begs the question whether there are other forms of angular weighting with greater SNRs while still being reasonable measures of differential phase contrast.  Whether the optimal weighting is sample specific is also unknown. That there is room for improvement is suggested by the slightly higher contrast transfer function obtained for bright-field ptycography compared to CoM imaging\cite{Pennycook2015}.
\section{Conclusion}
The advent of high-speed pixelated detectors has both motivated and aided a more detailed understanding of contrast in spatially-resolved diffraction patterns from non-periodic objects and the generation of images from diffraction data. Center-of-Mass imaging is one example of condensing the rich 4D diffraction data into a meaningful image.  In diffraction space, the scattering behavior and subsequent diffraction pattern formation was heavily dependent on the probe size versus feature size.  For potentials smaller than the probe size, an asymmetry appeared in the intensity within the bright field disk itself rather than the classically expected displacement.  The uniform displacement of the bright-field disk (and the full diffraction pattern) was instead a signature of a long-range potential with a uniform gradient.   By taking advantage of these differences in how long range and short range information are encoded, it is possible to distinguish slowly-varying magnetic domain contrasts from rapidly varying grain contrast by selecting the integration angles.  This approach works best at medium resolution where the probe is larger than structural features but smaller than the magnetic domains.

In real space, the CoM image is usually interpreted within the strong phase approximation as a convolution or correlation of the probe shape with the gradient of the potential.  However, for atomic potentials the CoM image profile is usually dominated by the shape of the probe rather than the sample potential given the strong singularity at the nucleus.  While the peak heights differed with the Z scaling of the atomic potential, the normalized shape profiles were very similar to the derivative of the probe shape, and subtle differences from outer shell electrons are likely to be much smaller than probe tails from residual geometric and chromatic aberrations.  Even peak deflections can be misleading; a small defocus of 1 nm led to an apparent change in atomic number of  ΔZ=19 for an isolated W atom.

The precise Z dependence of CoM is a function of convergence angles and beam voltage, reflecting the range of momenta sampled.  At smaller convergence angles and beam voltages, the CoM signal shows oscillations from the shell structure of the imaged atoms, while for larger cutoff angles, a smooth and closer-to-linear Z dependence is obtained.  As previously noted, this makes CoM and iCoM imaging well suited for images where both light and heavy atoms are present.  

We explored the effect of detector inner and outer angles on the SNR for DPC and CoM imaging of a single atom.  In both cases, the best SNR was achieved for an outer detector angle being larger than the bright field disk.  Overall, DPC showed a slightly higher SNR, and little sensitivity to increasing the outer angle.  In contrast, the CoM SNR drops once the outer angle is increased much beyond 1.5x the bright-field disk due to the angular weighting function enhancing regions with low signal.  In thicker samples, this angle might be pushed out by multiple scattering.  Conversely, the DPC SNR strongly decreased with increasing inner angle cutoff, but CoM did not, again consistent with their different angular weightings.  

Pixelated detectors and its applications are still in relative infancy.  While they allow a great deal of flexibility in reproducing traditional imaging modes in post-processing, their true strength may be characterization modes with non-trivial angular weightings and transformations, like CoM and ptychography, that are unique to pixelated detectors.  The rich contrast and signal diversity present in the recorded 4D data sets suggest rich opportunities in developing new imaging modes to exploit this new information.

\section{Funding}
This work was supported by the Air Force Office of Scientific Research through the 2D Electronics MURI grant FA9550-16-1-0031 (MC) and the National Science Foundation (NSF) through the Platform for the Accelerated Realization, Analysis, and Discovery of Interface Materials (PARADIM; DMR-1539918) (ZC). Support for the MM-PAD development was provided by the U.S. Department of Energy, grant DE-FG02-10ER46693.  The adaptation to the STEM was supported by the Kavli Institute at Cornell for Nanoscale Science.  Electron microscope and facility support from the Cornell Center for Materials Research, through the National Science Foundation MRSEC program, award \#DMR 1719875.
\section{Acknowledgements}
The authors acknowledge microscopy support from John Grazul and Mariena Silvestry Ramos.  The WSe2 sample was provided by Ming-Yang Li from the Lain-Jong Li group at King Abdullah University of Science and Technology.  We thank Mark Tate, Prafull Purohit, and Sol Gruner for help with the pixel array detector.
\bibliography{BibLibrary}
\end{document}